\documentclass[a4paper,11pt]{article}
\usepackage{jinstpub} 
\usepackage{todonotes}
\usepackage{doi}

\title{\boldmath Data acquisition from high-rate detectors at MAX~IV}

\author[a,1]{P. Bell,\note{Corresponding author.}}
\author[a]{M. Cascella,}
\author[b]{F. Engelmann,}
\author[a]{T. Eriksson,}
\author[a]{A. Lilius,}
\author[a]{Z. Matej,}
\author[a]{J. Metz,}
\author[a]{A. Salnikov,}
\author[c]{C. Weninger,}
\author[a]{and M. Yazdi}

\affiliation[a]{MAX IV Laboratory, Lund University,\\
Fotongatan 2, Lund, Sweden}
\affiliation[b]{Previously at MAX IV Laboratory, now Institute for Cybersecurity and Digital Trust, \\
The Ohio State University, 2015 Neil Ave, Columbus, USA}
\affiliation[c]{Previously at MAX IV Laboratory}

\emailAdd{paul.bell@maxiv.lu.se}

\abstract{At MAX~IV pixelated area detectors are operated at high frame rates to take advantage of the X-ray beam properties available from the fourth generation synchrotron in scattering, diffraction and imaging applications. A variety of photon counting and charge integrating detectors and sCMOS cameras have been integrated into a common data acquisition (DAQ) system in which data are streamed to a central Kubernetes cluster, mounting an IBM Storage Scale (GPFS) file system. The DAQ system provides live feedback from the detectors/cameras and extends to enable on-the-fly data processing.  Control system integration via Tango provides a standardised single interface for controlling all DAQ components. Starting from an overview of the detector types in use, we describe the design and implementation of the MAX~IV detector-DAQ system and report a quantitative study of its performance in terms of data throughput and detector operating rates.}


\begin{document}
\maketitle
\flushbottom

\section{Introduction}
\label{sec:intro}

The MAX~IV synchrotron laboratory in Lund, Sweden, is currently operating 16 beamlines for users, 
providing X-ray techniques in diffraction and scattering, imaging, and spectroscopy, covering a wide-ranging
scientific programme. 
The 3~GeV ring~\cite{machine} was the first fourth generation light source when it commenced operation in 2016, offering low horizontal emittance, high brilliance and a high degree of coherence.  
The ability to fully exploit these photon beam properties is dependent on
the performance of the detectors at the beamline experimental end-stations.
While each technique has its own specific requirements, there is a general need for
large area pixelated detectors to be run at high frame rates.
Such devices generate large volumes of data at high bandwidths which must be captured and stored
without losses, in experiments where the detectors are operated continuously for extended
periods. Moreover, it is not sufficient to simply record this data to disk for later analysis; to guide the experiments, on-the-fly analysis and visualisation must be provided.
These requirements motivate a combined data acquisition (DAQ) system and online analysis framework based on {\emph {data streaming}}.
The design, implementation and control of the DAQ system developed at MAX~IV for this purpose is described in this paper. We begin with an overview of the
detectors and cameras that are supported and conclude with a quantitative evaluation of the system performance.

\section{Detector overview}
\label{sec:detectors}

In general, the high brilliance of the 3~GeV ring permits the same statistical precision of a measurement to be achieved with reduced counting (exposure) times and hence motivates the use of high frame rate detectors, thereby increasing the throughput of any beamline technique. Hybrid photon counting detectors are the "workhorse" for many applications, being high frame rate while also offering zero noise through thresholding and high dynamic range. For a full discussion of their benefits see~\cite{hpc}. Photon counting detectors are therefore found at several beamlines on the 3~GeV ring, for imaging and diffraction/scattering techniques.

Photon counting detectors have certain limitations, however, and cannot satisfy all scientific requirements. First, the maximum photon counting rate is generally limited to around $10^7$~counts/s/pixel. For this reason the Macromolecular Crystallography (MX) beamline MicroMAX - with a flux at the sample of up to $10^{15}$ photons/s - employs a JUNGFRAU charge integrating detector from PSI~\cite{jf} in addition to a photon counting detector. Similarly, the FemtoMAX beamline at the Short Pulse Facility (SPF) operates with ultra short (100~fs) pulses for which a custom Time over Threshold (ToT) detector was developed, which can resolve of order 1\,000 photons per such pulse per pixel~\cite{femto}.  
A second limitation of photon counting detectors is the relatively large pixel size. Some techniques such as full field tomography, available at the ForMAX and DanMAX beamlines, require higher resolutions, with pixel sizes of the order of 10~$\mu$m; in these cases CMOS cameras are preferred. These typically operate in the visible light regime and are combined with a scintillator and focusing optics. CMOS cameras, and charge-coupled devices (CCDs), are also found at the SoftiMAX beamline, a soft X-ray beamline on the 3~GeV ring. Here it is the low photon energy in the range between 275 eV and 2.5 keV which reflects a third limitation, that photon counting electronics equipped with standard Silicon sensors are unable to detect a photon signal of less than a few keV, while CMOS sensors may have a usable quantum efficiency in that window.

A list of pixel area detectors on the 3~GeV ring - plus FemtoMAX at the SPF - is given in table~\ref{tab1}.
They are commercial detector systems from companies such as Dectris and XSpectrum and commercial scientific CMOS (sCMOS) cameras from Andor and Hamamatsu, among others. 
Some have been custom made by the vendor such as the four module "windmill" XSpectrum Lambda detector at ForMAX~\cite{formax} and the L-shaped Pilatus3 at CoSAXS, both designed to allow a combined SAXS-WAXS geometry.   
Only those which are integrated into the MAX~IV standard streaming DAQ system which this paper describes are included, so the 9~Megapixel (9M) and the 16M Eiger2 detectors at MicroMAX and BioMAX are not discussed further - these make use of the Dectris file-writer application adhering to the NeXus "MX Gold Standard" format~\cite{nxmx}.
The integration of the JUNGFRAU detector at MicroMAX is described in~\cite{jfmaxiv}. 
Other special cases include a Photron Nova S16 for high speed tomography which stores images to local memory and is only partially integrated into the streaming DAQ system.

Table~\ref{tab1} reports the detector size and typical maximum operating rate in frames-per-second (fps), from which the uncompressed data rate can be computed. The use of lossless compression significantly reduces the actual bandwidth usage for photon counting detectors. The stated operating frame rate may be less than what the detector is capable of for several reasons, such as the nature of the application, the simultaneous use of other slower measurement equipment or a limitation on the speed at which the motors of the experimental setup can be scanned. Hence the table is not intended as a comparison of the detector capabilities but a description of how they are currently used at MAX~IV.

\begin{table}[htbp!]
\caption{Currently operated pixel area detectors and their typical maximum frame rates in frames-per-second (fps), which may be less than the maximum capability of the detector, from selected beamlines on the 3~GeV ring and FemtoMAX. Corresponding {\emph{uncompressed}} data rates are indicated but for photon counting (PC) detectors a lossless compression is always applied which can reduce the rate by up to two orders of magnitude depending on the application (see section 3.1). Systems at MAX~IV on loan are not included.}
\centering
\label{tab1}
\smallskip
\begin{tabular}{l|lllll}
\hline
Beamline & Detector & Type & Pixels & Frame rate   & Data rate \\
          &  make \& model   &     &   approx   & fps & GB/s\\
\hline
NanoMAX & Dectris Eiger2 &PC & 500k & 100  & 0.2\\
        & Dectris Eiger2 &PC & 1M & 100  & 0.4\\
        & Dectris Eiger2  &PC & 4M & 100  & 1.6\\
        & Dectris Pilatus3  &PC & 1M & 25  & 0.1\\
CoSAXS  & Dectris Eiger2 &PC & 4M & 500  & 4.0\\
        & Dectris Pilatus3 "L-shaped" &PC & 2M & 100 & 0.8\\
ForMAX  & Dectris Eiger2 &PC & 4M & 500  & 4.0\\
        & XSpectrum Lambda "windmill" &PC & 3M & 500 & 4.5\\
        & Hamamatsu Orca Lightning &sCMOS & 12M & 120 & 2.0\\
        & Andor Zyla ($\times$3) &sCMOS & 5.5M & 100 & 0.8\\
DanMAX  & Dectris Pilatus3 &PC & 2M & 250  & 2.0\\
        & Hamamatsu Orca Lightning &sCMOS & 12M & 120 & 2.0\\
        & Andor Zyla &sCMOS & 5.5M & 100 & 0.8\\
        & XIMEA MX1510MR-SY-X4G3 &sCMOS & 150M & 6 & 1.4\\
Balder  &  Dectris Eiger ($\times$2) &PC & 1M & 500  & 1.0\\
SoftiMAX& Andor Zyla &sCMOS & 5.5M & 100 & 0.8\\
(soft X-rays) & Tucsen Dhyana &sCMOS & 4M & 24  & 0.2\\
        & Teledyne PI-MTE3 &CCD & 16M & 0.06  & 0.004\\
FemtoMAX &  Dectris Pilatus3 "ToT" &PC & 1M & 10  & 0.04\\
(SPF)    & Andor Zyla (numerous) &sCMOS & 5.5M & 10  & 0.08\\
         & Andor Balor &sCMOS & 17M & 10 & 0.25\\
\hline
\end{tabular}
\end{table}

\section{MAX~IV detector-DAQ system overview}
\label{sec:daq_intro}

\begin{figure}
    \centering
    \includegraphics[width=0.9\linewidth]{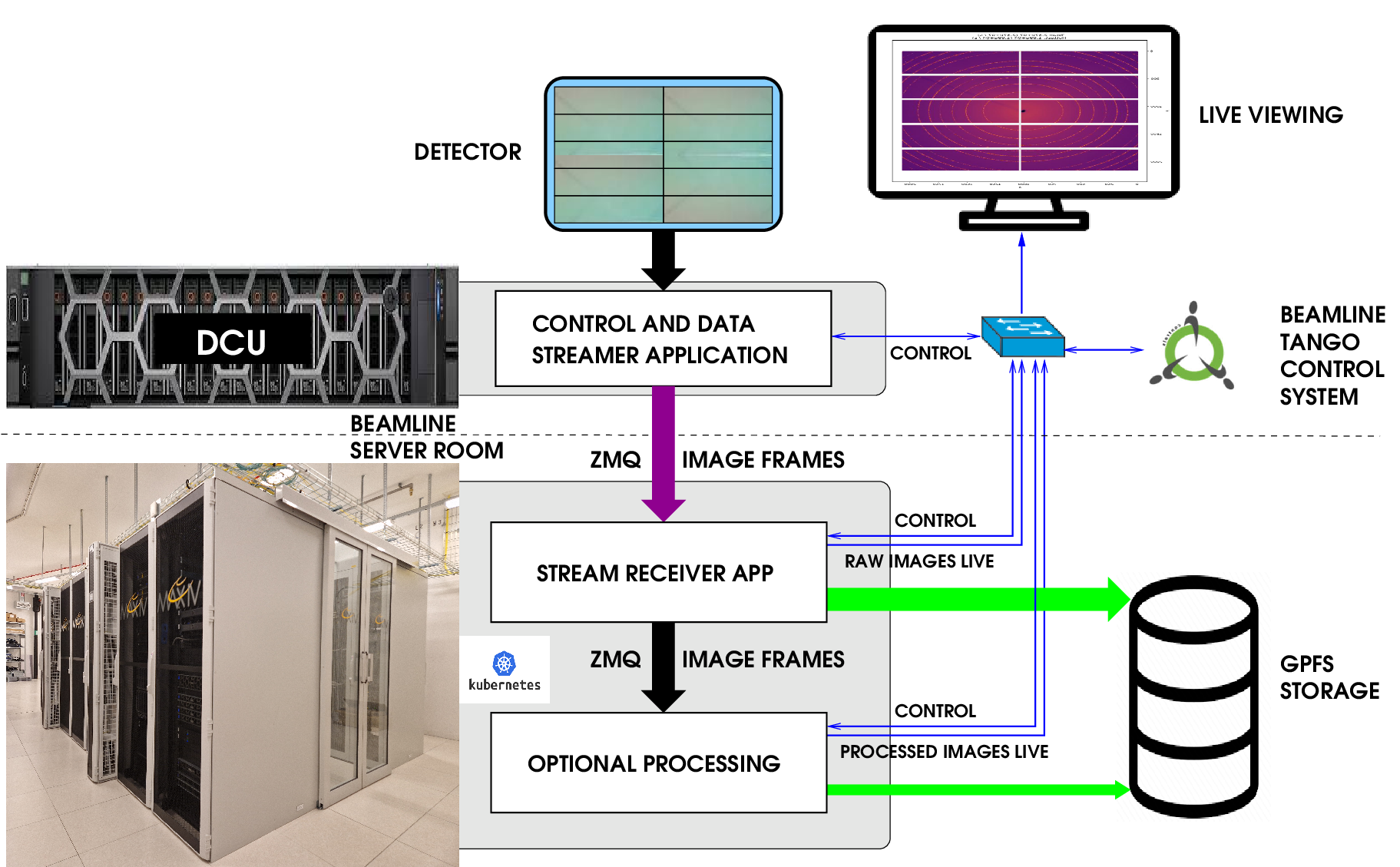}
\caption{Overview of the MAX~IV detector-DAQ scheme. A detector-specific data streamer application runs on the Detector Control Unit (DCU) which has a 40~Gb/s fast data connection (purple) to the Kubernetes DAQ cluster and a 10~Gb/s slow control connection (blue) to the beamline control network. A general purpose stream-receiver application runs on the DAQ cluster, with optional additional processing steps. Data are streamed between each layer over ZeroMQ. A further slow connection between the DAQ cluster and the beamline control network allows control and monitoring of the stream-receiver and for
accessing raw or processed images at low frame rates for live-viewing. Within the DAQ cluster there is a 100~Gb/s Ethernet network for intra-node traffic separated from the data ingest. I/O to the GPFS storage happens over dedicated Infiniband fabric (green). }
\label{fig1}
\end{figure}

An overview of the detector-DAQ scheme including a simplified view of the network connectivity is shown in figure~\ref{fig1}. 
For genuinely on-the-fly analysis and visualisation it is necessary to operate on a data stream rather than on files and ZeroMQ~\cite{zeromq} is used for this streaming of data (image frames).
The system is divided into a \emph{control and data streaming} layer, which is detector-specific and runs close to the detector at the beamline, and a data \emph{stream-receiver and processing} pipeline layer which is detector agnostic and takes place in a central Kubernetes-managed DAQ cluster mounting the IBM Storage Scale (GPFS) storage infrastructure. 
Data are saved in HDF5 files, directly as raw detector images and optionally after some processing step(s). 

The separation of potentially unstable detector systems at the beamline, perhaps running prototype detector software, and the GPFS storage avoids risking the stability of the latter, which was a key 
factor in the overall system design.
The role of Kubernetes is to automate the deployment and management of the workloads in the DAQ cluster and ensure the environment is easily scalable and resilient to hardware or software faults. 
Workload scheduling is based on resource demands, current load, availability and even the location of the origin of the data stream. Assignment of various DAQ workload properties, such as beamline-specific user IDs and storage locations, API URLs and network addresses is done at the infrastructure layer by leveraging the  capabilities of Kubernetes. The effort invested into the initial setup of the system has proven to have simplified and streamlined the resulting operational experience.

\subsection{Data streaming from the beamline}
\label{sec:daq_part1}

Hybrid photon counting detectors from commercial suppliers are usually delivered with a custom server referred to as the Detector Control Unit (DCU).  The communication protocol between the detector head and the DCU is typically a closed system over twisted pair or optical fibre.  In the case of CMOS cameras, MAX~IV prepares a server that fulfils the role of DCU. This may hold a "frame grabber" card supplied with the camera, for example a CoaXpress (4 Lane CXP-6) PCI Express card for the Andor Balor and Orca Lightning or a Camera Link (CL10) PCI Express card for the Andor Zylas. Some CMOS cameras such as the Tucsen Dhyana use USB~3.0. The distance between the DCU located in the beamline electronics rack and the  camera in the experimental hutch means that for these type of connections some method of range extension is often required, for example by conversion of the Camera Link or USB to optical fibre.

The Eiger detectors from Dectris~\cite{eiger} natively provide a stream of data from the DCU over ZeroMQ, typically over a single 40~Gb/s Ethernet interface. For all other detectors and cameras we  develop a streamer application that runs on the DCU
and generates a data stream from a ZeroMQ PUSH socket
adhering to a common format called STINS, which is similar to the Dectris Stream V1 format \cite{stream12}: a JSON-encoded start-of-series message, then each frame transmitted as a multipart message with a header holding some JSON meta-data followed by the image data blob (which may be in compressed format) and finally a JSON end-of-series message. A verifiable data model for the JSON part of the protocol is available in \cite{drsp_stins}.

The implementation of the streamer application is necessarily specific to the underlying interface provided for the particular detector or camera. Often, the streamer application can be developed and deployed on the DCU thanks to a high level API or SDK from the detector or camera supplier, which provides methods to get the image frames into memory. When possible, the streamer application is developed in Python, sometimes interfacing with the SDK/API  via the C Foreign Function Interface (CFFI). The Hamamatsu Orca Lightning and Andor cameras are such examples.

The Dectris Pilatus3~\cite{pilatus} and XSpectrum Lambda~\cite{lambda} detectors are special cases and merit further description. For the Pilatus, acquired images are written as files to a RAM disk on the DCU by the Dectris-installed "camserver" application.  Our streamer application is notified as these files are created, reads the image content into memory and then pushes it out over ZeroMQ. 
The XSpectrum Lambda is the only system currently in operation that is served by more than one DCU; in this case, one server per two of the four detector modules. 
The delivery of image frames into memory synchronised between the DCUs is handled by the manufacturer's SDK against which the streamer applications are built. Four streams are generated, one per module, which must be recombined in the receiver.  

For the photon counting detectors lossless compression is always applied on the DCU. In the case of the Eigers, Bitshuffle LZ4 (bslz4) is applied by the Dectris software before the frames are pushed. This can achieve a compression
factor ranging from a few to a few hundred depending on the application~\cite{bssaxs}. 
For the CMOS cameras, where the dark noise is not removed by the application of a signal threshold, the lossless compression does not work well and is not applied. This means that the highest full-frame data rate achieved today and handled by this standard DAQ system is from the 120~fps Hamamatsu Orca Lightning camera at around 2~GB/s. 

Most DCUs have a 10~Gb/s Ethernet connection to the beamline control system 
which at MAX~IV is based on Tango~\cite{tango}.
Except for the Eiger detectors, which have their own HTTP-based REST-like control interface, the developed streamer application also implements the control over the hardware, e.g. to set exposure times and start and stop acquisitions. 
Where we have full access to the DCU, as in the case of the CMOS cameras, we install Tango there and the streamer application is tightly coupled to the Tango device. Where this is not convenient, for example because the supplied DCU has a non-standard operating system\footnote{The MAX~IV control system standard operating system is Rocky Linux with some legacy CentOS machines}, a light streamer application is developed with the fewest possible dependencies. The application then provides its own remote interface to allow a Tango device (written in Python) running on one of the standard control system machines to interact with it. 

For all detectors and cameras, the Tango device presents a uniform interface, a MAX~IV standard detector device.  An essential, minimal set of commands and attributes are exposed, notably an Arm command and the self-explanatory attributes ExposureTime, TriggerMode and nTriggers, abstracting however these settings may be referred to in the underlying interface (SDK etc) and thereby hiding the complexity of the specific hardware. This makes the integration with any client software, 
such as Sardana~\cite{sardana} for scan orchestration, largely detector independent.

\subsection{Data reception in the DAQ cluster}
\label{sec:daq_part2}

On the DCUs the ZeroMQ stream is made available on a 40~Gb/s Ethernet interface (or 10~Gb/s if sufficient for the data rate of the detector).
Data (image frames) are then streamed to the DAQ cluster over dedicated 100~Gb/s Ethernet fabric, the so-called  "purple" network, in which direct long-range fibres connects the DCU to the network switch in one of the server halls housing the DAQ cluster. The purple network benefits from using so-called jumbo frames with 9\,000 bytes Maximum Transfer Unit (MTU) configured. This network is physically separate from the "blue" network of the control system, used for management traffic and live-viewing.

\subsubsection{IT infrastructure for DAQ}

\begin{figure}
    \centering
    \includegraphics[width=0.95\linewidth]{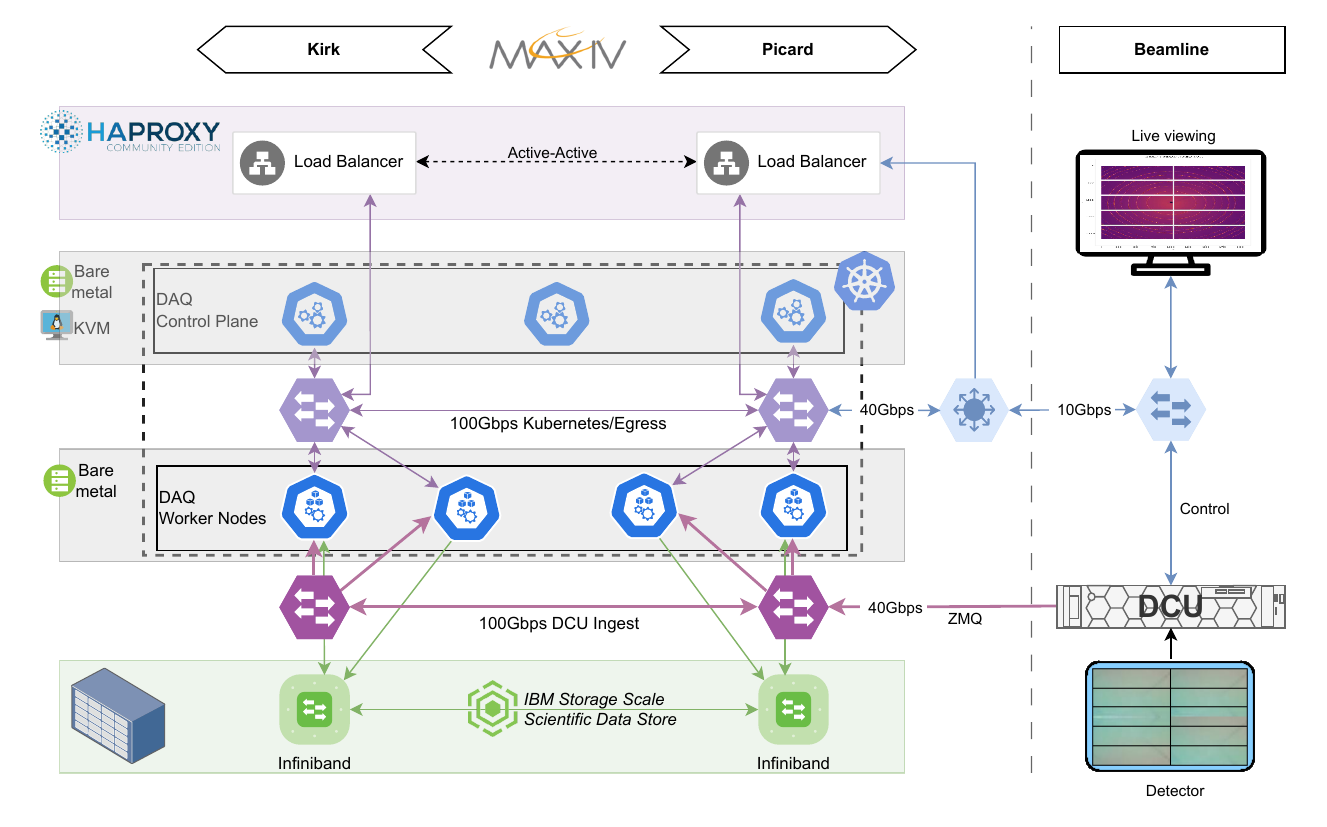}
    \label{fig:infra-overview}
\caption{Overview of the MAX~IV IT infrastructure for DAQ with a focus on the internals of the DAQ Kubernetes cluster. The DAQ Worker Nodes, Control Plane, Load Balancers and Storage Layer distributed over two server halls (\emph{Kirk} and \emph{Picard}) are interconnected via 100~Gb/s Ethernet and HDR Infiniband to receive, process and store data received via ZeroMQ over dedicated 100~Gb/s ingest Ethernet.}
\label{fig2}
\end{figure}

A schematic of the IT infrastructure layout that facilitates data reception in the DAQ cluster is depicted in figure~\ref{fig2}.
To enhance system availability, the equipment is evenly distributed across two server halls, designated \emph{Kirk} and \emph{Picard}, which are physically located in different parts of the building but share the same network through aggregated links. In addition to the nodes executing DAQ workloads, the cluster depends on common MAX~IV IT infrastructure services, such as load balancers.

The \emph{DAQ Worker Nodes} layer comprises the dedicated HPC nodes for running the stream-receivers and any further on-the-fly processing workloads.
At the time of writing, there are ten Dell PowerEdge R650 Worker Nodes, each with 2.9~GHz Intel\textregistered~Xeon\textregistered~Gold 6326 CPUs and 512~GB of RAM.
The traffic between nodes goes over 100~Gb/s Ethernet fabric, separated from the data ingest from the DCU. This network is also used to stream data out for live viewing and/or further processing in or outside the DAQ cluster. 
On the operating system level, the DAQ Worker Nodes are optimized for high-throughput I/O processing. 

The \emph{Control Plane} is a Kubernetes-specific layer running the cluster controllers, managing the nodes and configuration for running workloads. There are three KVM Virtual Manchines for the Control Plane function running alongside other VMs on the shared infrastructure. They have access to the same Kubernetes/Egress network segment, but no connections to the DCUs or storage fabric. 

On the top of the Control Plane, there is a common \emph{Load-Balancing} layer implementing the single entry points for managing the distributed system and terminating any HTTP traffic or control traffic in particular. This layer is not specific to the DAQ cluster, but the common IT infrastructure HAProxy setup.

Finally, the \emph{Storage Layer} 
based on the GPFS distributed file system
is natively mounted by the Worker Nodes over dedicated Infiniband fabric to avoid any interference with the data flows over Ethernet. 
The GPFS is implemented on two IBM Elastic Storage\textregistered~System (ESS) 3500 appliances to deliver high performance and scalability,
on which the partition for scientific data allocates 346~TB on the NVMe flash tier and
7.4~PB on the NL-SAS spinning disk tier.
Complementing this configuration, an IBM TS4500 Tape Library functions as the third, 10~PB cold storage tier.
The separation is transparent to the user\footnote{When the active storage reach 90\% capacity usage, the oldest data are migrated to tape until it reaches 70\% capacity, but are automatically recalled on to active storage on access.}. Currently, the annual data increase is around 3~PB/year.

\subsubsection{DAQ stream-receiver application}

\begin{figure}
    \centering
   \includegraphics[width=0.8\textwidth]{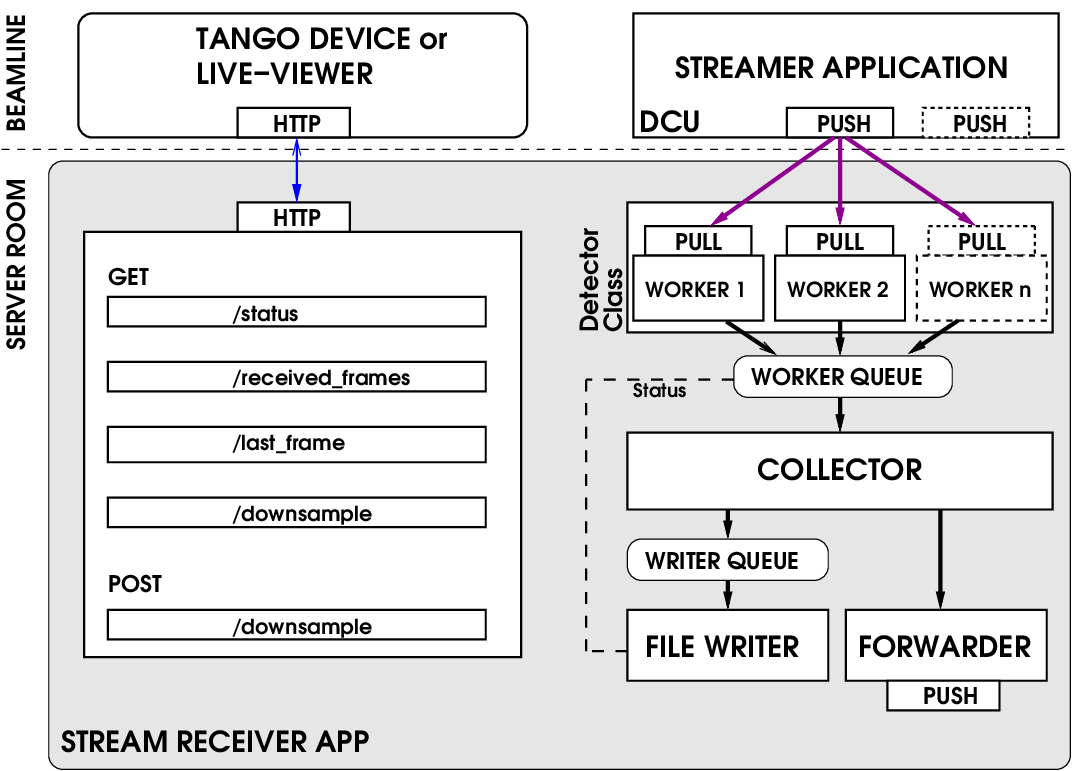}
\caption{Detail of the stream-receiver application which is deployed in the DAQ cluster as a Kubernetes pod for each detector. On the data handling side the ZeroMQ PULL sockets can be parallelised over multiple workers. The collector orders the frames before they are written to file or forwarded. On the API side an HTTP interface allows monitoring of the receiver status (e.g. by the Tango device for the corresponding detector) and access to the last frame (e.g. for the live-viewer) }
\label{fig3}
\end{figure}

Since the streamer applications running on the DCUs generates a similar data stream for all detectors and cameras, they are all handled by a common stream-receiver.
A schematic of this application - available at~\cite{ssmaxiv} - is shown in figure~\ref{fig3}. Written in Python, it consists of horizontally scalable number of \emph{workers} (one or many ZeroMQ PULL clients) to receive and apply basic transformations to the image frames, a \emph{collector} for ordering and distribution of the data to the HDF5 \emph{file writer}, and the \emph{forwarder} (ZeroMQ PUSH socket) which republishes the frames for additional processing in or outside the DAQ cluster.
The receiver software is designed to be easily extended and in addition to our STINS format (inspired by the Dectris Stream V1) already supports other stream standards including the Dectris Stream V2~\cite{stream2} and the SLS Detector Software~\cite{slsdetsw} 

The workers can pre-process the images they receive; they can rotate them, apply calibration corrections (e.g. the counts to energy calibration for the FemtoMAX ToT Pilatus) and handle decompression or compression operations. All these transformations are implemented either in C or Cython for performance reasons. For instance, for the Pilatus, the stream-receiver workers must uncompress the images recorded and sent from the DCU in Crystallographic Binary File (CBF) format and recompress them with bslz4 to be consistent with other photon counting detectors; we have experimentally determined that with eight workers, we can keep up with the 250~fps frame rate of the Pilatus3 2M detectors.

The collector uses the frame number included in the message headers to order them;  it is essential that the position of the frame data in the eventual HDF5 file matches the trigger number that generated it, so that it can be matched to other data from the same trigger in subsequent analysis. The worker and writer queues are used to pass both received frames and control messages; for instance, the file writer puts its status in the worker queue for the collector to process. The file writer process uses the h5py library and implements a minimal version of the NeXus NXDetector base class. The header message includes various meta-data fields of this NeXus class including detector settings such as whether various corrections are applied, the exposure time, etc.

The stream-receiver application provides an API through an HTTP interface for checking, for example, the state and status and number of received frames, and to obtain the most recent frame for live-viewing. The state and status follow a state machine with the following allowed states:
  \begin{itemize}
  \itemsep0em 
      \item
      Idle: when waiting for a header message indicating that a series of frames will follow;
    \item
    Running: entered after receiving the header message. Transitions back to Idle on receiving an end-of-series message;
    \item
    Error: for all unforeseen occurrences.
\end{itemize} 
This interface and state machine allows the stream-receiver application to be fully handled by the Tango device for the detector. In addition, a simple raw data live-viewer based on SILX~\cite{silx} is provided for all detectors and cameras. It works by making HTTP requests at 1~Hz to the stream-receiver for the latest frame and displaying it in a SILX GUI widget. The stream-receiver has an option to downsample the images for the live-view which is essential for some larger area cameras for which the data rate, even at 1~fps, would saturate the bandwidth of a client computer in the control room. 

Finally, data forwarded by the stream-receiver at full rate is the starting point for any following optional processing steps in the DAQ pipeline. 
Commonly used for many beamlines and techniques is a fast, parallelised azimuthal integration algorithm
using the "azint" library~\cite{pyazint} based on~\cite{matazint}. 
In a similar way to the stream-receiver, the azint pipeline also provides an HTTP interface
to allow control and visualisation of the processed data at the beamline. This type of online visualisation is an essential tool for the experimenter to judge the scientific quality of the data, thereby improving the efficiency of the beamtime and minimising collection of unwanted data.

\subsubsection{Stream-receiver instances on DAQ cluster}

Running the stream-receiver applications in the Kubernetes-managed DAQ cluster enables scalable and reliable deployment through containerized instances (\emph{pods}). The DAQ cluster infrastructure complemented by a \emph{daq-deployer} Helm Chart~\cite{helm} provides the foundation for deploying, monitoring and maintaining the applications effectively in the Kubernetes ecosystem.

\begin{figure}
    \centering
    \includegraphics[width=0.95\linewidth]{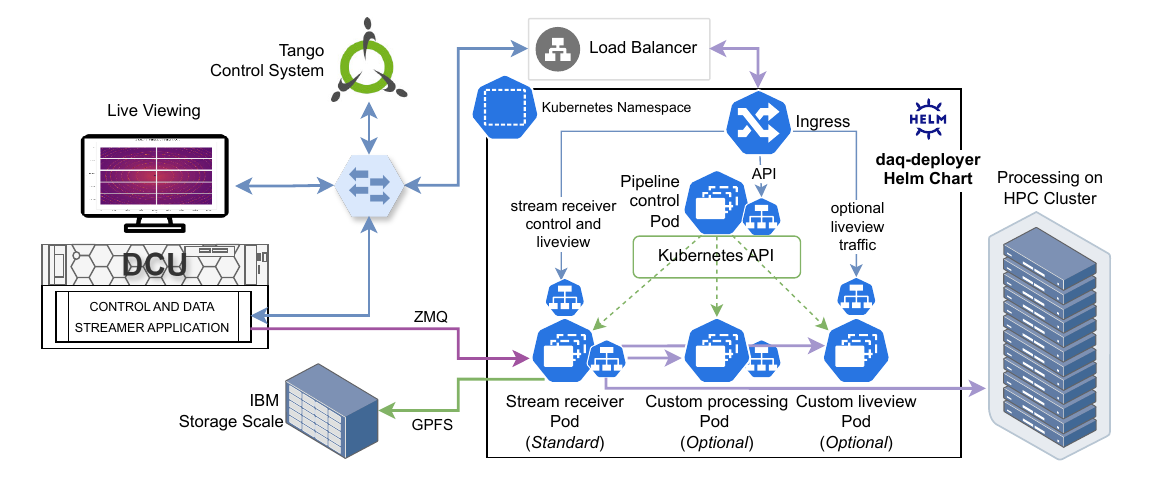}
    \label{fig:k8s-pods-overview}
\caption{Overview of the stream-receiver deployment with a focus on the Kubernetes resources created in the DAQ cluster by the \emph{daq-deployer} Helm Chart. The stream-receiver application packaged into a container image runs in a pod with a dedicated network interface to the particular detector or camera (purple). Stream re-publishing allows the pipeline to be complemented with optional processing and/or additional beamline-specific live-view steps (in addition to the raw-image live-view always available from the stream-receiver). The pipeline control pod provides the HTTP API to manage other pipeline pods. Ingress defines HTTP URLs accessible via the Load Balancer from the control system networks (blue) to allow API and additional custom live-view access from the beamline systems.}
\label{fig4}
\end{figure}

Figure~\ref{fig4} illustrates the Kubernetes workload structure for a single instance of a stream-receiver-based DAQ pipeline. All workloads configured for a specific detector or camera system are deployed into the dedicated Kubernetes \emph{namespace}. The namespaces of different detectors on the same beamline are grouped into common \emph{projects}\footnote{Project is a concept of Rancher~\cite{rancher} that is used to manage the DAQ cluster. They are technically the shared labels on the group of namespaces.}. 
On the infrastructure side, there are policies\footnote{Policies are implemented using the Kyverno~\cite{kyverno} policy engine as part of the DAQ cluster setup.} configured for each project to enforce the user IDs of the processes running inside the pods, the storage areas that are allowed to be mounted, the network access controls, etc.

The stream-receiver application is packaged into a container image. To run the instance of a container image for a particular detector or camera, the set of Kubernetes objects must be defined, including \emph{Deployments}, \emph{Services} and \emph{Ingresses}.
While it is possible to define these manually, the \emph{daq-deployer} Helm Chart provides better manageability and streamlines the user experience by templating the necessary objects based on a short YAML configuration file. Configuration is thus focused only on the key information (such as the IP address of DCU, the number of ZeroMQ worker PULL sockets to use, etc) instead of Kubernetes objects semantics.

The Kubernetes pod running the stream-receiver for a particular detector or camera has a dedicated SR-IOV network interface in the high-speed streaming network (purple). The beamline-specific storage area in the GPFS is mounted inside the pod, accessible over Infiniband. The processes run with a beamline service account which has permission to write data to that mounted directory. Once deployed, the HTTP API of the stream-receiver is exposed via the Kubernetes \emph{Ingress} accessible through the Load Balancer endpoint from the control system network (blue), allowing the Tango device to control both the detector hardware (via the streaming and control layer on the DCU) and the stream-receiver application.

In addition to the stream-receiver pod (containing the stream-receiver application) the pipeline created by the \emph{daq-deployer} instantiates the pods for any optional processing steps, such as "azint", and an overall pipeline control pod.
For the processing steps of the pipeline, the Kubernetes \emph{Service} for re-publishing the received ZeroMQ stream is defined. Depending of the type of the service, both internal republishing within the DAQ cluster (to processing pods) and external republishing outside the cluster (e.g. to stream data to a separate HPC farm for resource-intensive processing) are supported.
The pipeline control pod contains a service to manage Kubernetes deployments (e.g. starting/stopping the stream-receiver) from the Tango device or other clients that are not Kubernetes-aware.
It is defined by \emph{daq-deployer} as a separate deployment reachable via \emph{Ingress} (hence the blue control system network). The service exposes an HTTP API and translates the calls to actions against the Kubernetes API. For example, on receiving the stop command for the stream-receiver it would downscale the stream-receiver deployment. The pipeline control also makes possible the downscaling of idling workloads to improve the resource management.

The first version of the Helm Chart templated all DAQ cluster implementation-specific details, encapsulating infrastructure-specific knowledge. For example, the SR-IOV interface for the stream-receiver pod was templated into the \emph{Deployment} specification, hardcoding interface names and what IP address management configuration is used. 
This approach meant that any changes on the infrastructure side needed to be reflected in the Helm Chart and all workloads redeployed,
which did not scale well with an increasing number of deployments
and required awareness of all infrastructure details.
The current version of the \emph{daq-deployer} Helm Chart is built upon the policy engine. The infrastructure-specific knowledge is moved to the policies level, maintained separately as part of DAQ cluster setup. Policies mutate the Kubernetes objects' specifications based on an agreed set of labels. For example, defining the \texttt{daq.network.interface/purple:dhcp} label will trigger the defined rules in the policy engine to add all necessary SR-IOV specific configuration on pod creation. This separates infrastructure maintenance from the stream-receiver lifecycle (the primary concern of the detector software expert) and allows easy manual deployment of workloads for rapid prototyping. The \emph{daq-deployer} Helm Chart can focus on actual DAQ pipeline composition instead of hard-coding infrastructure details.


Each pipeline instance is deployed in a "GitOps" way, with the configuration YAML file for the \emph{daq-deployer} Helm Chart being stored in a deployment-specific git repository (specific to a certain detector at a certain beamline). This makes the creation, configuration management and operation of the instances of the stream-receiver based pipelines declarative, versioned and easy to maintain without any specific knowledge of the underlying infrastructure or Kubernetes in particular.

\subsection{System configuration and operation}

In continuous or "fly" scanning, which is the dominant mode of high rate operation, the detectors are externally triggered while at the same time one or more parameters (e.g. the sample position) are scanned and recorded in coincidence with the detector triggers. The state machine of the standard detector Tango device enforces that once the hardware is Armed to expect a certain number of triggers it remains in Running state until all those frames have been captured.  

The procedure for capturing a series of image frames involves a closed-loop communication between the Tango device and the stream-receiver in the DAQ cluster. The user configures parameters such as the destination filename and exposure time before Arming the device, which sends a header message to the receiver. On reception of the header message, the stream-receiver opens the file - as specified in the header - for writing and transitions into a Running state. Once both the detector and the receiver are ready, the Tango device also enters in Running state, signalling that the detector can be triggered. Each frame is then sent to the receiver until the pre-defined number of triggers is reached, at which point the end-of-series message is sent. On receiving this message, the receiver closes the file and reverts to an Idle state, only after which can the Tango device exit the Running state.

\section{System performance}

The ultimate performance of the DAQ system can be limited by both hardware and software components. For example, the streamer and receiver software built on ZeroMQ is subject to certain limitations imposed by the underlying architecture of that library. These limitations may in turn depend on the hardware (CPU, memory, network) on which the ZeroMQ-based applications run. Similarly, the ultimate speed of the HDF5 writer component of the receiver is dependent on the software implementation but ultimately on the performance of the hardware of the cluster nodes. For a system consisting of many interdependent components the entire chain must be tested as a whole to find the overall effective limit, and then the bottlenecks identified. 

To evaluate the operational capacity of the system, we have conducted tests in an isolated environment made to replicate the production system.
A data (image) stream generator takes the role of a real detector DCU, paired with an instance of the standard stream-receiver (with one or more workers). These two components were deployed on separate nodes within the DAQ cluster, using the 100~Gb/s ingest network interfaces for both streaming and receiving. 
The data generator mimics a real detector by generating dummy frames and pushing them using the STINS protocol 
to the stream-receiver, which receives them and writes them to disk in the usual way.
The generator's frame rate and frame size can be regulated,
allowing us to simulate potential future scenarios involving next generation detectors with significantly higher data rates than we see today,
identifying the limits of the current system.

\subsection{Observed performance limits of complete system}

The frame rate and size of the generated frames were adjusted to evaluate the point at which the stream-receiver (with one worker) is no longer able to keep up with the incoming data rate. 
Two aspects were evaluated: the maximum frame rate, in fps, from the generator to the stream-receiver and written to disk; and the maximum network throughput in GB/s, measured using the psutil
Python package during the tests. 
To cover a range of possible detector image sizes and compression rates, frame sizes of 20~kB, 200~kB, 2\,000~kB, and 20\,000~kB were used in the measurements. 
Every measurement was repeated ten times under identical conditions and the data points in all plots show the mean and standard error of the result.

\begin{figure}[p] 
    \centering
   \includegraphics[width=1\textwidth]{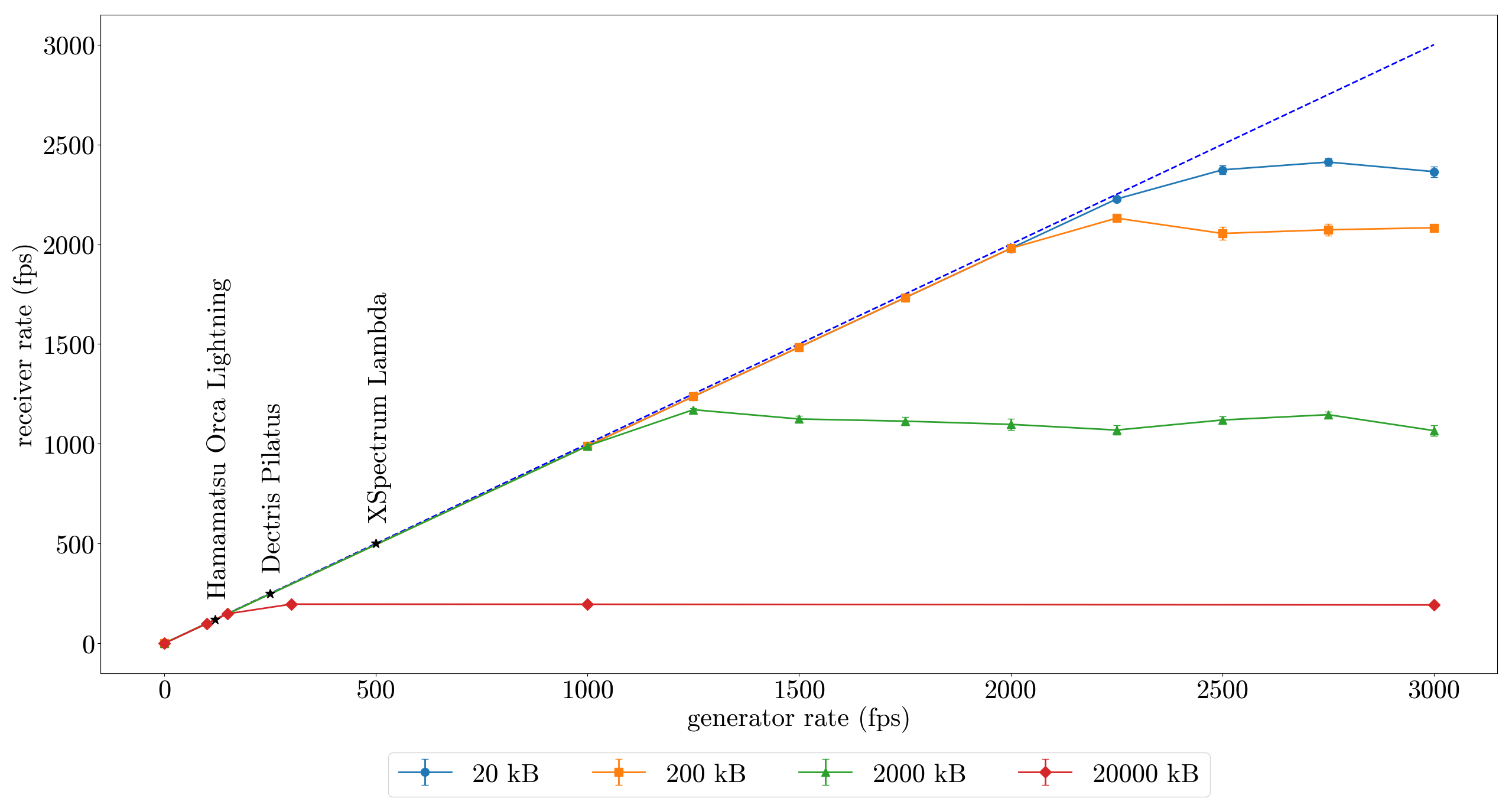}
\caption{Measured operating frame rate of a stream-receiver instance deployed in the MAX~IV DAQ cluster, including writing of data to disk, as a function of generated input frame rate, for different frame sizes. The linear dotted blue line shows the hypothetical performance where the stream-receiver keeps up with any generated frame rate. The Hamamatsu Orca Lightning (12M), Dectris Pilatus (2M), XSpectrum Lambda (3M) detectors have been marked on this line at their respective operating frame rates to provide perspective on the current uses of the MAX~IV system. }
\label{fig5}
\end{figure}

\begin{figure}[p] 
    \centering
   \includegraphics[width=1\textwidth]{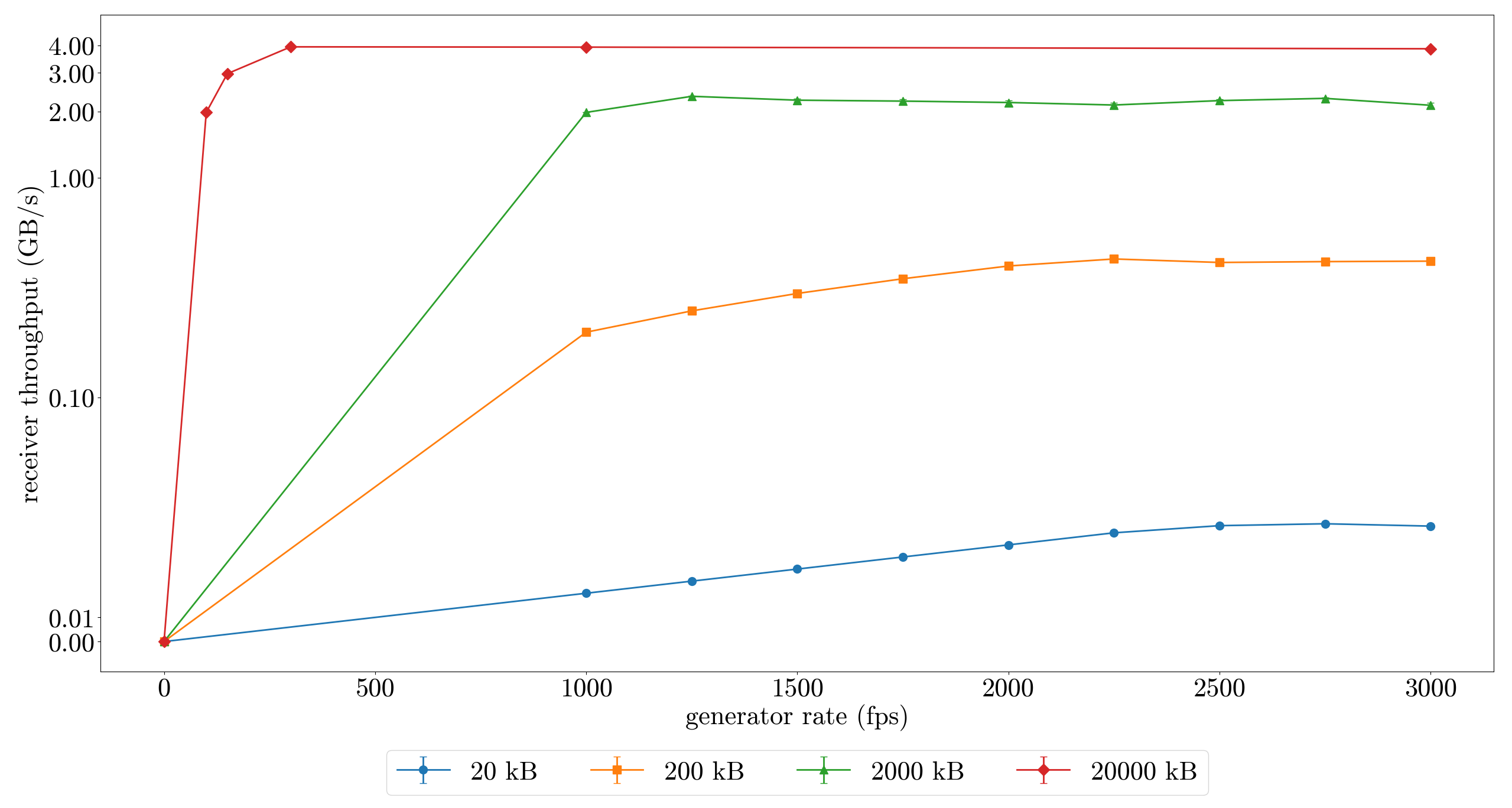}
\caption{Measured operating data rate of a stream-receiver instance deployed in the MAX~IV DAQ cluster, including writing of data to disk, as a function of generated input frame rate, for different frame sizes.}
\label{fig6}
\end{figure}

The results of the frame rate evaluation are shown in figure~\ref{fig5}, illustrating how well the stream-receiver keeps up with the generator's frame rate for different frame sizes. 
It can be seen that the highest frame rates that can be 
handled by the DAQ system are attained with the smallest frame sizes, 
for example around 2\,000~fps for frames of 200~kB.
Figure~\ref{fig6} however shows that the highest data throughput, of around 4~GB/s, is achieved with the largest frame size (20\,000~kB) at much lower frame rates.
This indicates a limit in the DAQ chain in operations per second rather than a network bandwidth restriction. Small frames are not processed proportionally faster than large ones, i.e. 20\,000kb frames can be processed at 200~fps but 200~kb frames cannot be processed at 20\,000~fps and instead use a much smaller fraction of what is shown to be the available data throughput of at least 4~GB/s.

\subsection{Performance optimisation of data streaming}

\begin{figure}[p] 
    \centering
   \includegraphics[width=1\textwidth]{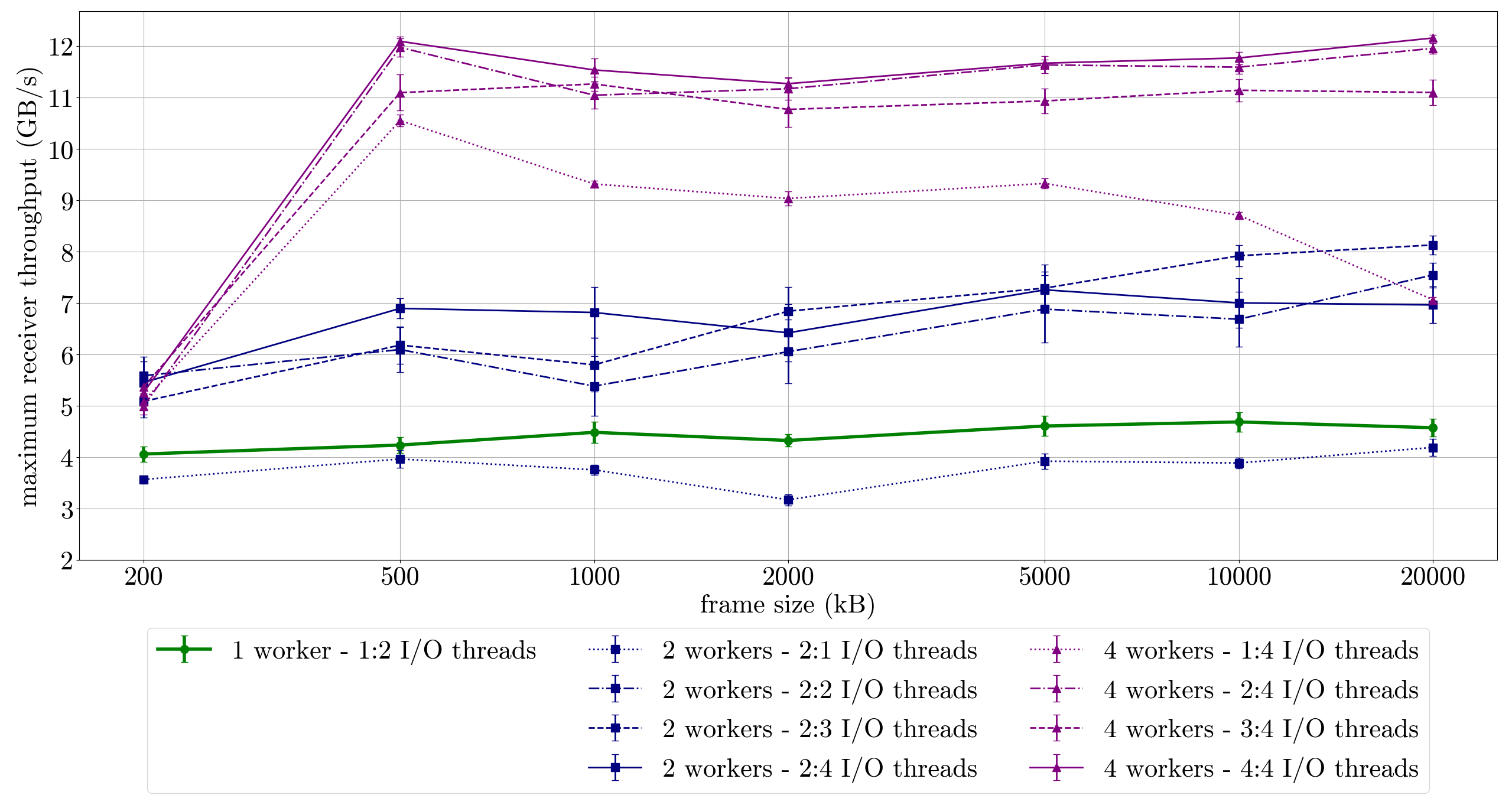}
\caption{Measured maximum operating data rate of a stream-receiver instance deployed in the MAX~IV DAQ cluster, excluding writing of data to disk, as a function of generated input frame size, for 1, 2 and 4 workers (ZeroMQ PULL sockets) in the stream-receiver application and with a selection of different ZeroMQ I/O context thread ratios.
}
\label{fig7}
\end{figure}

Studying the system as a whole, the origins of the above frame or data rate limits are not known. However, one explanation can be the writing of files to disk with the h5py library. If the file writing is disabled, the stream-receiver (still with one worker) {\emph{is}} now observed to run at over 20\,000~fps for frames down to 200~kb and thereby make full use of the 4~GB/s bandwidth.
This is illustrated by the thick line in figure~\ref{fig7} which now shows the maximum data throughput for different frame sizes.

\begin{figure}[p] 
    \centering
   \includegraphics[width=1\textwidth]{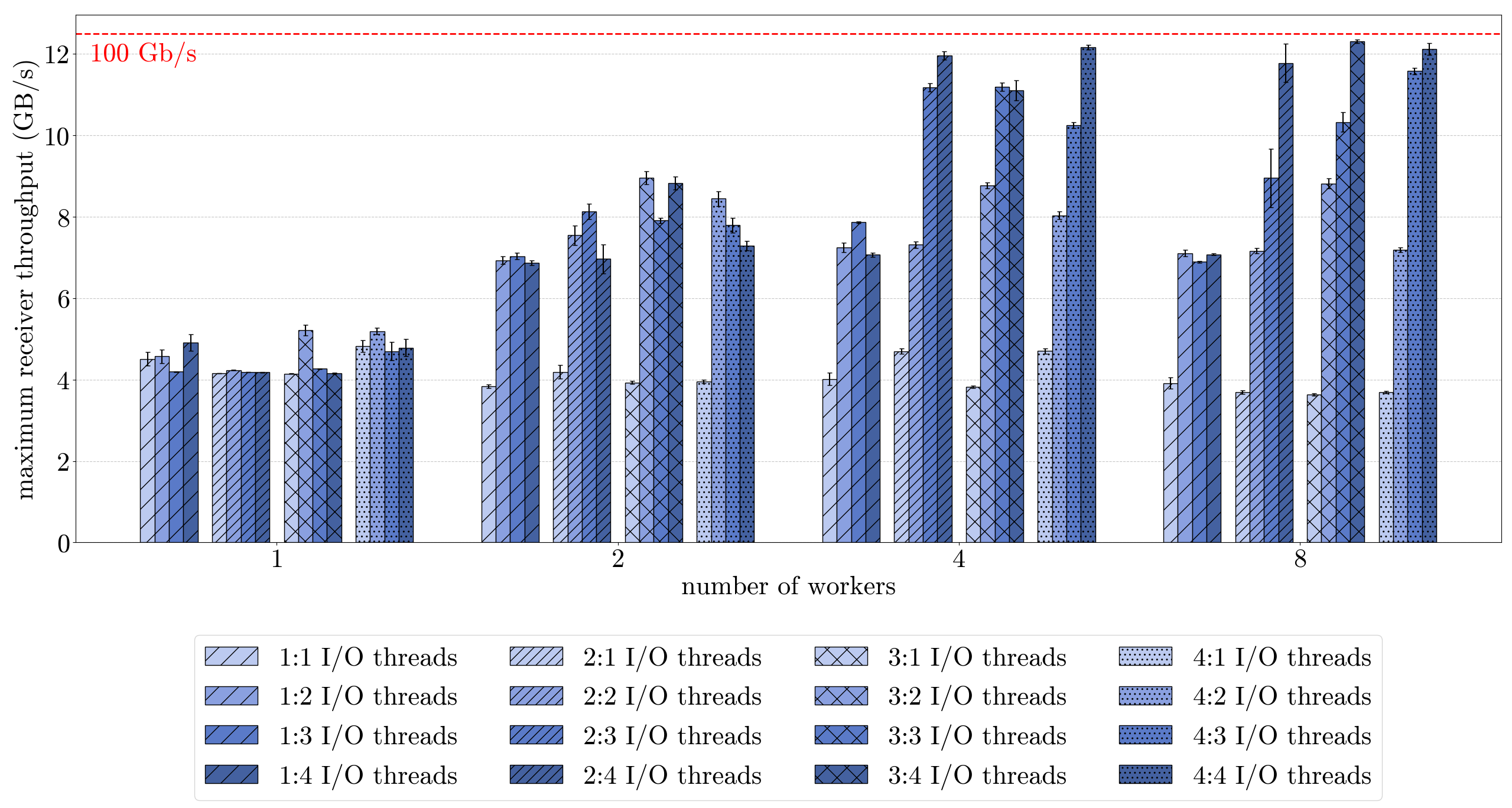}
\caption{Measured maximum operating data rate (using 20\,000~kB frames) of a stream-receiver instance deployed in the MAX~IV DAQ cluster, excluding writing of data to disk, for 1, 2, 4 and 8 workers (ZeroMQ PULL sockets) in the stream-receiver application and a range of ZeroMQ I/O context thread ratios.}
\label{fig8}
\end{figure}

Figure~\ref{fig7} furthermore highlights two additional factors that significantly influence the throughput of the stream-receiver application. First, the stream-receiver has an adjustable number of worker threads (ZeroMQ PULL sockets) assigned to each instance deployed on the DAQ cluster. Increasing the number of workers raises the maximum achievable throughput, up to a point. Second, the ZeroMQ library uses internal I/O threads that determine how much data a single ZeroMQ context can process and how many concurrent connections it can sustain. In contrast, the number of ZeroMQ sockets does not directly affect throughput; sockets act as logical endpoints that all share the same context and utilize the same I/O thread pool. Therefore, tuning the number of context I/O threads is essential for achieving maximum network performance. This applies to both sending and receiving sides — if only one side is properly configured, the system will saturate unevenly. 

One worker thread with one ZeroMQ context I/O sending thread and two I/O receiving threads is the production configuration. 
Figure~\ref{fig7} shows the maximum data rates (without file writing) for selected other combinations of number of stream-receiver workers and number of ZeroMQ internal I/O threads For the largest frame sizes, doubling from one to two workers allows the 4~GB/s data rate to be doubled, as expected, if the I/O threads are also increased. For smaller frames there remains a slightly lower limit but the dependency on frame size is much reduced now file writing is disabled. With four workers and appropriate I/O threads the 100~Gb/s network limit is reached for large frames.

As figure~\ref{fig7} suggests, the effects of the number of stream-receiver workers and ZeroMQ I/O context threads are interdependent. This is further studied in figure~\ref{fig8} where now the frame size is fixed at 20\,000~kB and the maximum data throughput found for a more exhaustive combination of numbers of stream-receiver workers and I/O threads. When only a single worker is used the maximum throughput is limited by however many I/O threads are configured. Increasing the number of workers without increasing the number of I/O threads leads to saturation in the I/O layer, which is most evident in the configuration with eight workers in figure \ref{fig8}. Here, increasing the I/O thread ratios from 1:1, 2:2, 3:3 to 4:4 produces an almost linear increase in throughput.
It can be concluded that with appropriate settings the current MAX~IV DAQ system will be able to handle data rates from a single DCU up to the full capacity of the 100~Gb/s networking.

\section{Summary and future improvements}

At MAX~IV we have created a DAQ architecture for high frame and data rate detectors based on data streaming over ZeroMQ. Instances of a standardised stream-receiver are deployed in a central Kubernetes DAQ cluster, which mounts a high performance GPFS data storage, and receive data from the detector DCUs at the beamlines over dedicated 100~Gb/s Ethernet fabric. 
The standard detector Tango device provides a simplified user interface to any detector or camera type - including control over the data stream-receiver in the DAQ cluster - so that non-experts need only deal with common basic attributes and commands. Integration to higher level SCADA systems such as Sardana for scanning orchestration is also simplified. 

Today there are over 60 stream-receiver instances deployed in the Kubernetes DAQ cluster, covering the detectors and cameras in table~\ref{tab1} plus others. 
In terms of "DevOps", having a standard approach simplifies maintenance and operation and the work needed to integrate a new detector is confined to the development of the streamer application on the DCU which is unavoidably detector specific. The addition of a new detector on the data receiving side requires only a new deployment of a stream-receiver in Kubernetes, by means of preparing the simple YAML configuration file for the \emph{daq-deployer} Helm Chart in a git repository. 

The current full data streaming and file writing paradigm can accommodate detectors generating up to 2\,000~fps, though the interplay between frame rate and data rate limits favours large frame sizes at small frame rates to use the most of the available data bandwidth. With a single ZeroMQ worker node, frame sizes of 200~kb, corresponding to a 
1M sized photon counting detector with 16 bit image depth and a factor 10 bslz4 compression, can be handled at over 2\,000~fps (limited by the frame rate), whereas frames of sizes 20\,000~kb can be handled at around 200~fps  (which is proportionally faster and now limited by the data throughput). These limits are compatible with the current operational needs in which high frame rate (1\,000~fps range) photon counting detectors have modest frame sizes thanks to the built-in compression and lower rate (100~fps range) CMOS cameras deliver high frame sizes due to their smaller pixels and lack of compression.

To keep pace with the increasing frame and data rates expected from future detectors, further development of the DAQ system will be necessary.
Improving the performance of the file writer or parallelising the file writing would significantly enhance the overall throughput of the stream-receiver. 
With that bottleneck addressed we have shown that with correct scaling of worker nodes in the stream-receiver and appropriate tuning of the ZeroMQ I/O threads we can saturate the full capacity of the 100~Gb/s networking between the DCUs at the beamline and the DAQ cluster.

The MAX~IV DAQ system, in addition to its core role of recording detector images to disk with 100~\% reliability, i.e. no frame loss, 
is intrinsically designed to provide a means of pseudo-real-time online data processing and visualisation, which will be described in detail in a future paper. 
Multimodal data pipelines based on the streams of data from multiple sources are under development.
As detectors become faster and data rates grow we foresee a greater need for this type of online analysis in the future, in order to address the "data deluge" on the end user, which will require a tighter coupling between the detector DAQ and on-the-fly data processing.

\end{document}